\documentclass[lettersize,journal]{IEEEtran}
\usepackage{amsmath,amsfonts,amssymb}
\usepackage{algorithmic}
\usepackage{array}
\usepackage[caption=false,font=normalsize,labelfont=sf,textfont=sf]{subfig}
\usepackage{textcomp}
\usepackage{stfloats}
\usepackage{url}
\usepackage{verbatim}
\usepackage{graphicx}
\def\BibTeX{{\rm B\kern-.05em{\sc i\kern-.025em b}\kern-.08em
    T\kern-.1667em\lower.7ex\hbox{E}\kern-.125emX}}
\usepackage{balance}
\usepackage{multirow}
\usepackage{float}

\usepackage{xcolor}
\usepackage{color,soul}

\usepackage{caption}
\usepackage{colortbl}
\usepackage{hyperref}
\usepackage{booktabs}
\usepackage[ruled]{algorithm2e}
\SetKwComment{Comment}{$\blacktriangleright$}{}
\SetCommentSty{normalsize}
\newcommand{\ali}[2]{\makebox[#1][l]{#2}} 

\begin{document}
\title{Example-Based Framework for Perceptually Guided Audio Texture Generation}

\author{
    \IEEEauthorblockN{Purnima Kamath\IEEEauthorrefmark{1}, Chitralekha Gupta\IEEEauthorrefmark{1}\textit{ Member IEEE}, Lonce Wyse\IEEEauthorrefmark{2}\textit{ Member IEEE}, Suranga Nanayakkara\IEEEauthorrefmark{1}}
    \\\IEEEauthorblockA{
        \IEEEauthorrefmark{2}Universitat Pompeu Fabra
    }
    \\\IEEEauthorblockA{
        \IEEEauthorrefmark{1}National University of Singapore
        \\\textit{Corresponding Author}: purnima.kamath@u.nus.edu
    }
    
}

\markboth{}%
{Example-Based Framework for Perceptually Guided Audio Texture Generation}

\maketitle

\begin{abstract}
Controllable generation in StyleGANs is usually achieved by training the model using labeled data. For audio textures, however, there is currently a lack of large semantically labeled datasets. Therefore, to control generation, we develop a method for semantic control over an unconditionally trained StyleGAN in the absence of such labeled datasets. In this paper, we propose an example-based framework to determine guidance vectors for audio texture generation based on user-defined semantic attributes. Our approach leverages the semantically disentangled latent space of an unconditionally trained StyleGAN. By using a few synthetic examples to indicate the presence or absence of a semantic attribute, we infer the guidance vectors in the latent space of the StyleGAN to control that attribute during generation. Our results show that our framework can find user-defined and perceptually relevant guidance vectors for controllable generation for audio textures. Furthermore, we demonstrate an application of our framework to other tasks, such as selective semantic attribute transfer.
\end{abstract}


\section{Introduction}
Audio textures are sounds generated by the super-position of multiple similar acoustic events~\cite{mcdermott2011sound,saint2021analysis}, such as the sounds made by water filling a container or a wooden drumstick repeatedly hitting a metal surface. Guided or controllable generation of such sounds using deep neural networks is useful for generating background environmental sound scores for movies, games, and automated Foley sound synthesis~\cite{anderson1997sound, moffat2019sound}. Such guidance during generation is usually achieved by conditioning generative models using semantically labeled data. For instance, impact sound textures can be semantically guided using object or material properties of the impact surface and a continuously varying water-filling texture can be guided using attributes such as the fill level of the container. While large datasets for audio textures can be readily recorded, labeling these sounds using semantic attributes such as material hardness or fill level is difficult. Therefore, to control generation, we develop a method to infer the vectors for semantic attribute guidance without the supervision of large labeled datasets.

Generative adversarial networks (GANs)~\cite{goodfellow2016nips} such as StyleGANs~\cite{karras2020analyzing,karras2019style} generate semantically disentangled latent spaces by learning the most statistically significant factors of variation within a dataset. Such disentangled latent spaces can be analyzed to find guidance vectors for controllable generation. We define semantic attributes in audio as
a set of factors that matter to human perception of sound~\cite{watcharasupat2021controllable}. Thus, to control the generation of audio textures, we analyze the disentangled latent space of a StyleGAN, to find guidance vectors based on user-defined semantic attributes.

Recent research has focused on describing semantic attributes to guide audio generation using text or language abstractions~\cite{kreuk2022audiogen, borsos2022audiolm, agostinelli2023musiclm,liu2023audioldm}. This multi-modal guidance is achieved by training generative models on large audio datasets in conjunction with text annotations such as captions~\cite{elizalde2022clap,guzhov2022audioclip}. While abstracting control using such language-based prompts is a step towards generalized sound generation, the research on their ability to granularly control audio texture synthesis (i.e. controlling the low-level auditory aspects of the sound) is still underway.

In this paper,  we propose using audio examples to guide latent space access and navigation. Similar to music information retrieval (MIR) techniques such as query-by-example~\cite{grosche2012audio,foote1997content,wang2006shazam,birmingham2001musart} and query-by-humming~\cite{ghias1995query, cartwright2014synthassist,kim2019improving,zhang2018siamese} we generate synthetic sound examples representative of the semantic attribute we want to control during generation.  We encode these examples into the latent space of a StyleGAN unconditionally trained on real-world audio textures. Then we use these latent embeddings to define guidance vectors in the latent space along which desired semantic attributes can be systematically varied during texture generation. As shown 
on our webpage\footnote{\href{https://pkamath2.github.io/audio-guided-generation}{https://pkamath2.github.io/audio-guided-generation}}, we use these guidance vectors to guide texture generation for various user-defined semantic attributes such as ``Brightness'', ``Rate'' or ``Impact Type'' for impact sounds and ``Fill-Level'' for the continuously varying texture of water filling.

We validate the effectiveness of our method for user-defined semantic guidance of texture generation through a comprehensive attribute rescoring analysis. We also conduct perceptual listening tests to evaluate the effectiveness of our method in changing specific attributes for various randomly generated sounds. In summary, our contributions are: 

\begin{itemize}
    \item An \underline{E}xample-\underline{B}ased \underline{F}ramework (EBF) to find user-defined attribute guidance vectors to semantically control audio texture generation.
    \item A synthetic audio query approach for latent space exploration of a generative model.
    \item An application of our framework for the task of semantic attribute transfer between textures.
\end{itemize}

\section{Related Work}

\subsection{Supervised Controllability in Audio}
Generative models for music, such as~\cite{engel2017neural, engel2019gansynth}, 
enable controllability by training on datasets with labels. This supervision helps organize the model's latent space according to the timbre-specific features in the datasets. Musical instrument datasets are usually labeled during dataset creation~\cite{engel2017neural}, and such labels are used to conditionally train generative models using attributes for pitch, loudness, or instrument timbres~\cite{engel2019gansynth,lee2018conditional}. 
Further, some architectures~\cite{nistal2020drumgan} additionally condition generation by extracting attributes such as sharpness or warmth automatically from the sound using feature extractors such as Audio Commons~\cite{audiocommons} and Essentia~\cite{essentia}. Similarly, DDSP~\cite{engel2020ddsp} based architectures, such as DDSP-SFX~\cite{liu2023ddsp}, extract attributes such as loudness and pitch from the sounds to condition generation. While such supervised training methods are highly effective for modeling musical instrument sounds, their effectiveness is limited in the domain of textures due to the lack of large-scale semantically labeled audio texture datasets. Further, the attributes used to control generation for inharmonic audio textures are different as compared to those of musical sounds~\cite{gaver1993we,gaver1993world}. For instance, when synthesizing impact sounds, we are more likely to be interested in controlling the object or material properties (such as impact surface hardness, etc.) than attributes such as pitch or loudness typically associated with musical sounds. Currently, there is a lack of audio texture datasets with such object or material property labels that can be used for supervised training.

To circumvent this lack of attribute labels for audio textures, MorphGAN~\cite{gupta2023morphgan} uses features extracted from the penultimate layer of a classifier for supervision to generate smooth texture morphs. Similarly, DarkGAN~\cite{nistal2021darkgan} is trained on soft labels distilled from an audio tagging classifier~\cite{kong2020panns} trained on tags from the AudioSet ontology~\cite{gemmeke2017audio}. The Sound Model Factory~\cite{wyse2022sound} trains a GAN which is used to create novel timbres followed by an RNN trained on sounds produced from the GAN and conditioned on points along smoothly parameterized trajectories through the GAN latent space. All of these supervised training methods rely on additional class or parametric information while training generative algorithms. Since GANs, particularly StyleGANs, can disentangle the latent space based on semantic attributes in the training data~\cite{karras2020analyzing}, our research explores finding user-defined semantic directions in the latent space of a StyleGAN to guide generation without the need for any explicit conditioning or labeled data during training. 

\subsection{Unsupervised Controllability}
In computer vision, algorithms such as~\cite{harkonen2020ganspace,shen2021closed} leverage StyleGAN's ability to disentangle the latent space to find directional vectors for editing semantics on images. Similarly, in audio, GANSpaceSynth~\cite{tahiroglu2021ganspacesynth} applies the GANSpace algorithm to control a pre-trained GANSynth trained on musical instruments in an unsupervised manner. More recently, in computer vision, Semantic Factorization (SeFa)~\cite{shen2021closed} performed better than other unsupervised algorithms to find vectors for controllable generation in the latent space of a pre-trained GAN. In this method, the weights of the layers that create the disentangled representation are decomposed to find the vectors for maximum variation. Such vectors are then used to edit semantics on unconditionally generated images. However, the directional vectors generated using SeFa need to be semantically labeled manually after observing edits across multiple samples. 

For speech and music~\cite{ggan_9492807, gaae_9272282} infer controllability based on supervision from a few labels. For images, FLAME~\cite{parihar2022everything} uses supervision from a few positive-negative image pairs by semantic editing and inverting real images in a StyleGAN's latent space. Direction vectors for semantic attribute editing are found by optimizing for cosine similarity between the pairs' difference vectors. In our work, we modify the FLAME method for audio textures and propose using a few fully synthetically generated examples to assist in deriving vectors in the latent space of a StyleGAN for attribute controllability. A cluster of similar synthesized audio examples is inverted~\cite{chai2021latent} to define clusters in StyleGAN's latent space. A prototype~\cite{snell2017prototypical} latent vector is derived from each cluster and is an abstract average of the semantic cluster they represent. Since such prototypes are designed to differ in a specific attribute, the difference vector between them in the latent space can be used for guiding audio texture synthesis and for semantic attribute transfer.

\subsection{Synthetic Texture Generation}
While real-world sounds could also be inverted to find latent representations in a trained StyleGAN, they are much more difficult to control than parametric acoustic sound synthesizers  ~\cite{SASPWEB2011,serra1990spectral,serra1990system,serra1997musical} or physics-based models~\cite{PASPWEB2010}. For our inharmonic textures, we use a physically informed synthesis technique found in William Gaver's seminal work on auditory perception~\cite{gaver1993world,gaver1993we}. His approach is based on the idea that humans hear and describe sound events in terms of their sources and source attributes better than in terms of the acoustic properties of the sounds themselves. The sound events in Gaver sounds are modeled on the physics of the objects interacting to produce the sound, such as the hardness of the material under impact or the force of impact. Gaver~\cite{gaver1993world} refers to analysis-by-synthesis as a process of updating synthesis parameters to match a target sound, which we use to discover StyleGAN latent vectors with synthetic audio queries. 

Although algorithmically synthesized sounds can sound unnatural, we employ them only for querying and searching the latent space of a StyleGAN. Multi-event synthetic textures can be quickly and easily generated using an analysis-by-synthesis approach with attributes adequate for this exploration task.

\section{Proposed Framework}\label{sec:method}
\begin{figure*}[t]
\centering
 \includegraphics[width=0.99\textwidth]{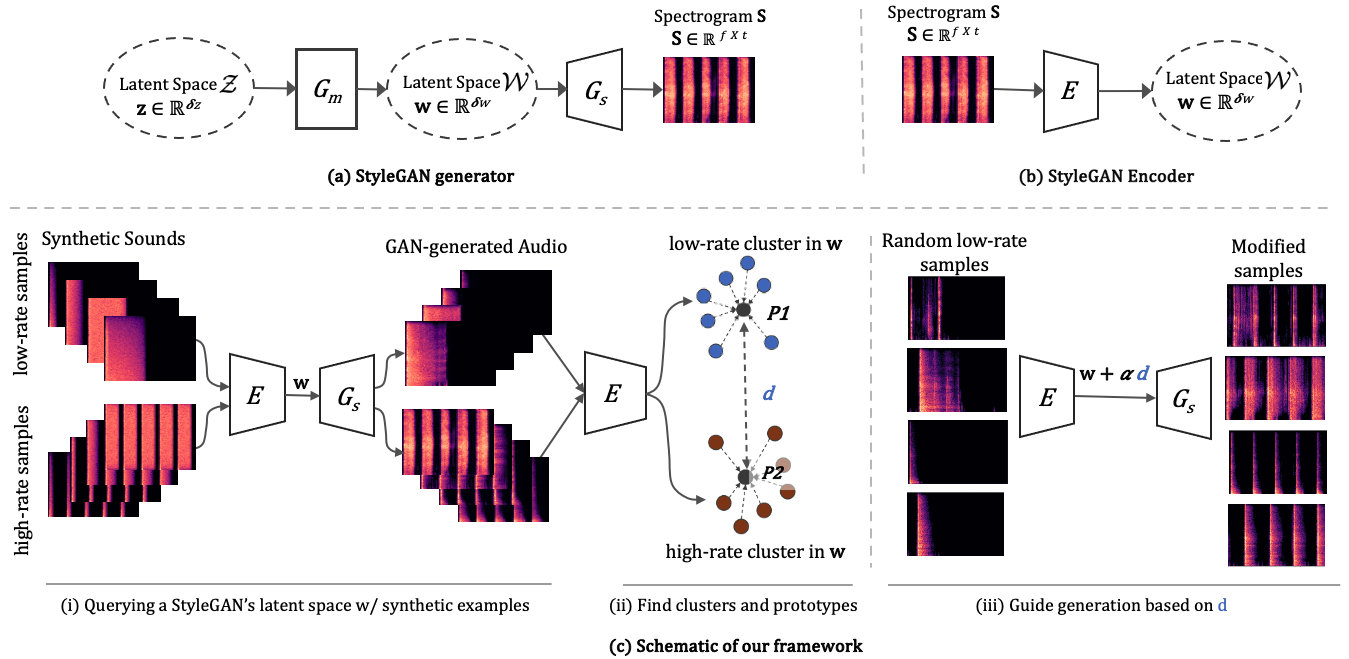}
 \caption{Schematic outlining the modules within our framework. (a) A StyleGAN's generator. Mapping network $G_m$ maps latent space $\mathcal{Z}$ to intermediate latent space $\mathcal{W}$ ($\mathbb{R}^{\delta_z}\rightarrow\mathbb{R}^{\delta_w}$). Synthesis network $G_s$ maps an intermediate latent vector $\mathbf{w}$ to spectrograms $\mathbf{S}$ ($\mathbb{R}^{\delta_w}\rightarrow\mathbb{R}^{f \times t}$). (b) Schematic of an Encoder $E$ which inverts spectrograms to the intermediate latent space $\mathcal{W}$ ($\mathbb{R}^{f \times t}\rightarrow\mathbb{R}^{\delta_w}$). (c) Schematic of our framework during inference. 
 }\label{fig:archdiag}
\end{figure*}

As shown in Figure~\ref{fig:archdiag}, we partition our goal to find semantic attribute vectors for controllable texture generation and propose a framework comprised of the following modules: 

\begin{itemize}
    \item A Generator module ({$G_s$}) of a StyleGAN trained on real-world audio for high-fidelity texture synthesis,
    \item A GAN Encoder ($E$), also known as a GAN inversion network, to encode an audio example into the latent space of a pre-trained StyleGAN,
    \item A parametric Gaver synthesizer for sounds used to locate desired points in the latent space of the StyleGAN,
    \item An algorithm to derive semantic attribute clusters and prototype vectors for guiding semantic synthesis trajectories in the latent space of the StyleGAN.
\end{itemize}
Figure~\ref{fig:archdiag}c illustrates our framework (during inference). $G_s$ is a StyleGAN generator and $E$ is the GAN Encoder. (i) We generate synthetic Gaver sounds for a semantic attribute we want to control. In the diagram above, we demonstrate this using ``Rate'', or the number of impact sounds in a sample, as the semantic attribute. We encode these synthetic sound examples into the latent space of a StyleGAN to find their $\mathbf{w}$ embeddings. (ii) Next, we derive the semantic attribute clusters and generate prototypes using the algorithm elaborated in section~\ref{subsec:semantic_directions}. The direction vector to guide generation for that semantic concept is indicated by ``$\mathbf{d}$''. (iii) Shows how we can use direction vector ``$\mathbf{d}$'' to guide generation on any randomly generated audio sample to increase or decrease ``Rate''.

\subsection{GAN for Audio Textures}\label{subsec:gan-arch}
While our framework can be applied to derive attribute guidance vectors within the latent space of any pre-trained generative model, such as Variational Autoencoders~\cite{esling2018generative}, Progressive GANs~\cite{nistal2020drumgan}, or StyleGANs for audio, 
in this paper, we demonstrate this using StyleGAN2~\cite{karras2020analyzing} trained on audio textures. Figure~\ref{fig:archdiag} (a) shows a schematic of a StyleGAN2's generator. We have excluded the discriminator section of StyleGAN2 in the schematic for brevity. Overall, a StyleGAN2's generator can be modeled as a function $G(.)$ that maps a latent space $\mathcal{Z}$, where $\mathbf{z}\in\mathbb{R}^{\delta_z}$, to the higher dimensional spectrogram space $\mathbf{S}\in\mathbb{R}^{f\times t}$, such that $\mathbf{S} = G(\mathbf{z})$. Here ${\delta_z}$ is the dimensionality of the $\mathcal{Z}$ space, and $f$, $t$ are the number of frequency channels and time frames of the generated spectrogram, respectively. StyleGANs further learn an intermediate representation $\mathcal{W}$, where $\mathbf{w}\in\mathbb{R}^{\delta_w}$, between that of $\mathcal{Z}$ and $\mathcal{S}$ via a mapping network $G_m(.)$. This intermediate latent space further disentangles factors of variation as compared to the latent $\mathcal{Z}$ space~\cite{karras2020analyzing}. Further, a synthesis network $G_s(.)$ maps the $\mathbf{w}$ vector to a spectrogram $\mathbf{S}$. A StyleGAN's intermediate $\mathcal{W}$ latent space is considered to be more disentangled, in terms of the various factors of variation in the training data, than its $\mathcal{Z}$ space~\cite{karras2019style}. We thus operate our framework and method in the intermediate latent space $\mathcal{W}$ to find semantically meaningful directions for controllability during generation.

\subsection{GAN Encoder}\label{subsec:gan-inversion}
Figure~\ref{fig:archdiag} (b) shows a schematic of our Encoder. While GANs learn to map latent space embeddings to real-world sounds, GAN inversion techniques learn inverse mapping, i.e., from the real-world sounds to the latent space embeddings. We adapt the encoder model from~\cite{chai2021latent} to estimate a $\mathbf{w}$ vector from an audio spectrogram randomly sampled from a pre-trained StyleGAN2. This model is based on the ResNet~\cite{he2016deep} architecture. Residual Network (or ResNet) architectures use stacks of residual blocks (a set of convolutional layers with skip connections) to learn residual functions with reference to the layer inputs. Such architectures have been previously successfully used for large-scale audio classification tasks~\cite{resnet_for_audio_classification}.

The input to the Encoder, as shown in Figure~\ref{fig:archdiag} (b), is a spectrogram of the audio sample to be inverted. Previously,~\cite{huang2022masked,niizumi2022masked} have shown that masking techniques for spectrograms are effective while learning generalized vector representations for audio. We extend this idea of arbitrarily masking the spectrogram to learn a $\mathbf{w}$ vector representation from the Encoder. This approach is especially useful during inference to generalize the encoder to synthetic Gaver sounds. It assists in projecting the synthetic sounds into a reasonable part of the latent space even though the encoder (or the GAN) is not directly trained on these sounds. Note that while training the Encoder, the weights of the StyleGAN2 generator are frozen. We only optimize the Encoder weights during training.

For noisy textures, such as the sounds made by water filling a container, we further employ amplitude thresholding of the spectrogram during training. This thresholding ensures that the encoder ignores the low-level noise and focuses on the most prominent events and frequencies in the spectrogram while estimating the $\mathbf{w}$ vector. To train the Encoder, we modify the loss function from~\cite{chai2021latent} to estimate only in the $\mathcal{W}$ space instead of $\mathcal{Z}$ as:

\begin{multline}\label{eq:loss_eq}
\mathcal{L} = \displaystyle{\mathbb{E}_{\mathbf{z}\sim{\mathcal{N}(0,1)}, \mathbf{w}=G_m(\mathbf{z}), \mathbf{S}=G_s(\mathbf{w})}[{\lVert \mathbf{S} - G_s(E(\mathbf{S})) \rVert}_2^2}\\
 \displaystyle{+ {\lVert \mathbf{w} - E(\mathbf{S}) \rVert}_2^2]}
\end{multline}

In Equation~\ref{eq:loss_eq},  $G_m(.)$ is the mapping network, $G_s(.)$ is the synthesis network of the StyleGAN2, and $E(.)$ is the Encoder that inverts a spectrogram $\mathbf{S}$ into the $\mathcal{W}$ space. While training the encoder, we randomly sample a  $\mathbf{z}$ from the $\mathcal{Z}$ space to generate the target spectrogram $\mathbf{S}$ using $G(\mathbf{z})$. We estimate the $\mathbf{w}$ for this spectrogram using the encoder $E(\mathbf{S})$. For the first loss term, we pass the inverted $\mathbf{w}$ through the synthesis network of the generator $G_s(\mathbf{w})$ and find the mean squared error (MSE) loss between the original and reconstructed samples. The second term is the MSE loss between the actual and the estimated $\mathbf{w}$ vector.

The loss function of the original Encoder algorithm~\cite{chai2021latent} additionally used a perceptual similarity loss term called LPIPS~\cite{zhang2018unreasonable} that calculates the distance between image patches to preserve the perceptual similarity of the estimated images. In our experiments, we evaluate the need for such perceptual loss terms for our task in comparison with our loss formulation in Equation~\ref{eq:loss_eq}.

\subsection{Synthesizing Gaver Sounds}\label{subsec:analysis-by-synthesis-approach}
To generate audio examples for querying the GAN latent space, we use two Gaver synthesis methods - (1) based on physical parameters of the interacting objects and (2)  based on object resonance as a series of bandpass filters. The first method is useful in generating sharp impact sounds or dripping sounds, and the second is for producing a larger variety of impacts and scraping sounds. More formally, a synthetic impact sound can be described as -
\begin{equation}
F(t) = \sum_n\phi_n {e}^{\zeta_n t}cos\omega_nt    
\end{equation}

\noindent where $F(t)$ describes the generated sound, $\phi_n$ is the amplitude of the $n^{th}$ partial, $\zeta_n$ is a damping constant, and $\omega$ is the frequency of the partial, $\sum$ signifies a sum over the total number of partials. From an ecological perspective, each component in the equation controls a physical aspect of the objects interacting to generate the impact sound. For instance, $\zeta$ in the equation controls the material hardness, $\phi$ controls the force of impact, and $\omega$ and $n$ control the size of the object. 

Method 2 creates impact and scraping sounds by passing Gaussian noise $\mathcal{N}(0, I)$ through band-pass and fade filters. The amplitude of the impact sound governs the force of impact, while the frequency bands, together with damping provided by linear or exponential fade filters, govern the material of impact of the sound.

\subsection{Semantic Clusters, Prototypes, and Guidance Vectors}\label{subsec:semantic_directions}
\begin{figure}[t]
\centering
         \includegraphics[width=0.3\textwidth]{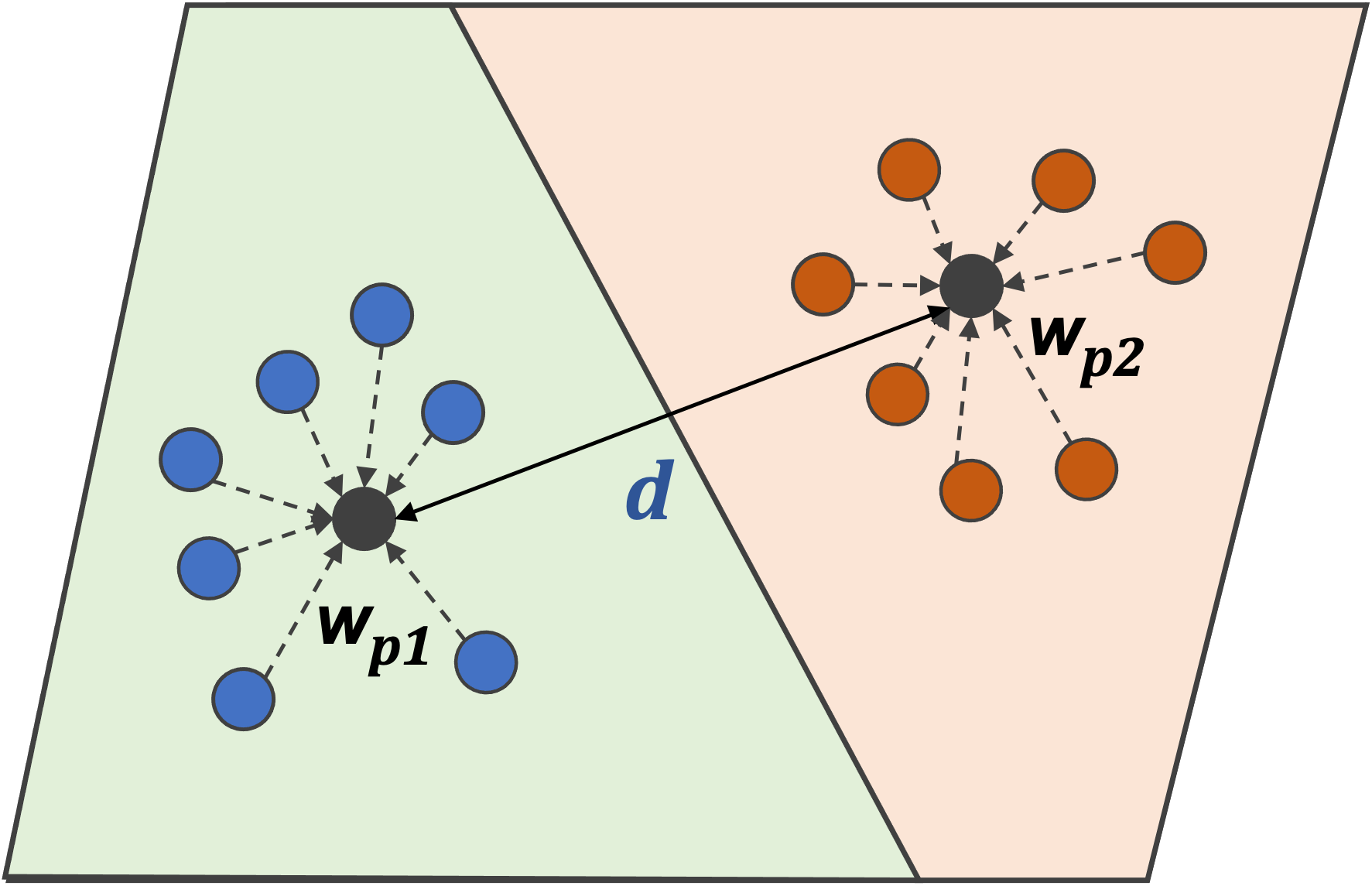}
         \caption{Schematic for generating semantic attribute clusters, prototypes $\mathbf{w_{p1}}$ and $\mathbf{w_{p2}}$, and the direction vector $\mathbf{d}$.}\label{fig:concept-prototypes-schematic-1}
\end{figure}

Having generated synthetic Gaver sounds, we invert them into the latent space of the StyleGAN to generate the $\mathbf{w}$ embeddings. We then cluster the sounds together in the $\mathcal{W}$ space to generate prototypes as shown in Figure~\ref{fig:concept-prototypes-schematic-1}. 

Assume, for example, that we want to derive directional vectors to control the attribute of ``Brightness'' of an impact sound. We define brightness as an attribute that indicates the presence or absence of high-frequency components in a sound. We generate a cluster of Gaver sounds where the semantic attribute is present (represented by blue dots in the figure), and another cluster of Gaver sounds where the semantic attribute is absent (or ``dull'' impact sounds represented by orange dots). We find the prototypes $\mathbf{w_{p1}}$ and $\mathbf{w_{p2}}$ representative of each semantic attribute cluster using Algorithm~\ref{alg:get_prototype}.

\IncMargin{1.5em}
\SetAlgoNoLine
\begin{algorithm}
    \caption{Get Prototype}\label{alg:get_prototype}
    \SetKwInput{KwData}{Input}
    \SetKwInput{KwResult}{Output}
    \SetKwProg{Fn}{Function}{:}{}
    \Indm\Indmm
    \KwData{\\
    \Indp
    $\mathbf{W_n}$ is a matrix of $\{\mathbf{w_0},...,\mathbf{w_n}\}$ encoded synthetic samples as column vectors, such that $\mathbf{W_n}\in\mathbb{R}^{\delta_w\times{n}}$\;
    $\mathbf{w\_avg}$ is a column vector for the center of mass of $\mathcal{W}$ space\;
    }
    \BlankLine
    \KwResult{\\
    \Indp
    $\mathbf{w\_ptype}$ the prototype representation\;
    }
    \BlankLine
    \Fn{$\mathnormal{GetPrototype}$ {($\mathbf{W_n}$, $\mathbf{w\_avg}$)}}{
        \ali{4.5em}{$\mathbf{W}\in\mathbb{R}^{\delta_w\times{n}} $} ${}\leftarrow \mathbf{W_n} - \mathbf{w\_avg}$\;
        \ali{3.5em}{} \Comment{Subtract $\mathbf{w\_avg}$ from each column of $\mathbf{W}$}
        \ali{4.5em}{$\mathbf{U},\mathbf{S},\mathbf{V}$} ${}\leftarrow \mathbf{SVD}(\mathbf{W})$\;
        \ali{4.5em}{$\mathbf{s}$} ${}\leftarrow diag(\mathbf{S})$\;
        \ali{4.5em}{} \Comment{Extract singular values from diagonal}
        \ali{4.5em}{} {matrix $\mathbf{S}$ as vector $\mathbf{s}$}\;
        \ali{4.5em}{$\mathbf{w\_ptype}$} ${}\leftarrow \mathbf{w\_avg}+{{\mathbf{u_s}}}{{{\mathbf{u_s}}}^T}{\overline{\mathbf{w}}}$\;
        \ali{4.5em}{} \Comment{$\overline{\mathbf{w}}$ is mean $\mathbf{w}$ sample vector from $\mathbf{W_n}$}
        \ali{4.5em}{} \Comment{${\mathbf{u_s}} \leftarrow {\mathbf{U}[:, argmax(\mathbf{s})]}$}
        \textbf{return} $\mathbf{w\_ptype}$
    }
    \BlankLine
\end{algorithm}
\DecMargin{1.5em}

To generate our prototypes, we adapt a technique from computer vision for generating Eigenfaces. First, we shift or center the inverted $\mathbf{w}$ embeddings of the synthetic samples by subtracting the center of mass of the $\mathcal{W}$ space, namely $\mathbf{w\_avg}$, from them. $\mathbf{w\_avg}$ is the the mean $\mathbf{w}$ vector encoded from our training set. These mean-subtracted $\mathbf{w}$ embeddings record how each synthetic sample differs or varies w.r.t the mean sample in the $\mathcal{W}$ space. Next, we stack all the mean-subtracted $\mathbf{w}$ embeddings for the synthetic samples in a semantic cluster together as columns of a matrix. We perform singular value decomposition on this matrix and select the component associated with the maximum singular value to construct the prototype. The intuition behind doing this is that after decomposition, the component with the highest singular value has the most common prominent feature amongst all the samples being analyzed, i.e., the semantic attribute being modeled. Furthermore, by modeling the mean-subtracted $\mathbf{w}$ embeddings, we ensure that we model the variations in the $\mathbf{w}$ vectors better instead of focusing on the shared common features encoded by $\mathbf{w\_avg}$. Constructing a prototype this way is more robust to outliers or artificial synthesis artifacts.

The difference between the $\mathbf{w}$ embeddings of the two prototypes $\mathbf{w_{p1}}$ and $\mathbf{w_{p2}}$, denoted as direction vector ($\mathbf{d}$), can be used to continuously and sequentially edit the semantic attribute as follows - 

\begin{multline}\label{eq:direction_edit}
\qquad\qquad\qquad\mathbf{w_{edited}} = \mathbf{w} + \alpha * \mathbf{d}
\\
\noindent{
(\textrm{where } 0 < \alpha < 1)
}
\end{multline}

\noindent where $\mathbf{w}$ is a randomly chosen $\mathcal{W}$ vector, `+' and `*' indicate element-wise operations, and $\alpha$ is a continuous scalar parameter that signifies step size. Larger values of $\alpha$ correspond to a greater degree of semantic attribute edit on the sample. Further, using $-\mathbf{d}$ reverses the direction of the edit. The edited sounds can be reconstructed by passing the $\mathbf{w_{edited}}$ through the StyleGAN2 synthesis network $G_s(.)$.

\section{Experiments}
\subsection{Datasets}\label{subsec:datasets}
We use two audio texture datasets in our experiments: (1) The Greatest Hits dataset~\cite{owens2016visually} to demonstrate the effectiveness of our approach on impact sounds and (2) a Water filling a container dataset~\cite{gupta2022parameter} for noisy audio textures. Through these two datasets, we demonstrate the effectiveness of our method to cover a range of event-based and noisy textures.

\subsubsection{The Greatest Hits Dataset}
This dataset contains audio and video recordings of a wooden drumstick probing indoor and outdoor environments by hitting, scraping, and poking different objects of different material densities. We use this dataset to explore the rich timbres arising out of the interactions between the wooden drumstick and various hard and soft surfaces such as tree trunks, dirt, leaves, metal cans, ceramic mugs, carpets, soft cushions, etc. The dataset contains approximately 10 hours of denoised audio split into 977 audio files each approximately 35 seconds. Each file contains impact sounds interacting with different types of objects. We split the audio files into consecutive 2-second sounds sampled at 16kHz to unconditionally train our StyleGAN2. We develop semantic attribute clusters, prototypes, and attribute guidance vectors for the attributes \textbf{\textit{Brightness}} (whether the sound contains mostly high-frequency components or is dark or dull containing mostly low-frequency components), \textbf{\textit{Rate}} (whether the number of impact sounds in a sample is high or low), and \textbf{\textit{Impact Type}} (whether the sounds are sharp impacts or scraping/scratchy sounds made by dragging the stick across the surface). 

\subsubsection{Water filling a container}\label{subsec:dataset-water}
This dataset~\cite{gupta2022parameter} contains 50 audio recordings of water filling a container at an approximately constant rate for an average duration of $\sim$30 seconds.  We develop semantic attribute clusters, prototypes, and attribute guidance vectors for the continuously varying attribute of \textbf{\textit{Fill-Level}} of the container. We sample the recorded audio files using a sliding window of 100ms to generate approximately 10,000 2-second audio files sampled at 16kHz to unconditionally train our StyleGAN2. We choose a small sliding window size of 100ms to achieve better interpolatability~\cite{watcharasupat2021controllable} for \textbf{\textit{Fill-Level}} in the $\mathcal{W}$ space of the GAN. 

\begin{figure*}[t]
\vspace{-15pt}
\centering
    \includegraphics[width=\textwidth]{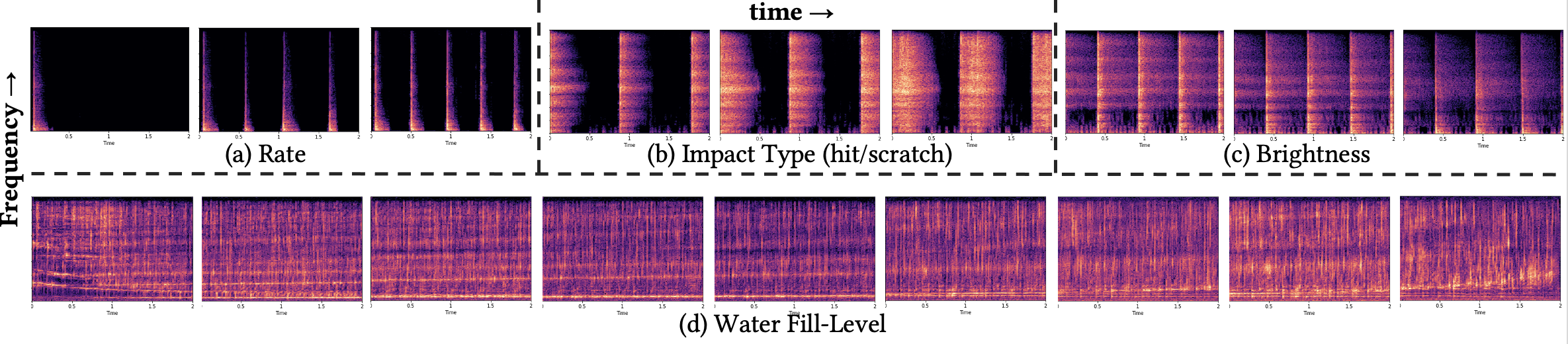}
    \caption{(Top Row) Spectrogram examples of guided generation using our method based on change in the attributes of (a) Rate (increases L to R), (b) Impact Type (becomes scratchy L to R), and (c) Brightness (decreases L to R). Note that for each example, as one attribute changes, the other attributes do not undergo a change. (Bottom Row) Examples of guided generation for water filling a container based on Fill-Level. Note how the Fill-Level and its respective frequency components gradually increase from L to R. All sounds can be auditioned on our webpage \url{https://pkamath2.github.io/audio-guided-generation}. 
    }
    \label{fig:teaser}
\vspace{-5pt}
\end{figure*}

\subsection{Implementation Details}\label{subsec:exp_setup}  
\quad\textit{StyleGAN2:} We set $\mathcal{Z}$ and $\mathcal{W}$ space dimensions $\delta_z$ and $\delta_w$ both to $128$ and use 4 mapping layers in the Generator for all our experiments. Further, we use the log-magnitude spectrogram representations generated using a Gabor transform~\cite{marafioti2019adversarial}(n\_frames$=256$, stft\_channels$=512$, hop\_size$=128$), a Short-Time Fourier Transform (STFT) with a Gaussian window, to train the StyleGAN2 and the Phase Gradient Heap Integration (PGHI)~\cite{pruuvsa2017noniterative} for high-fidelity spectrogram inversion of textures to audio~\cite{gupta2021signal}. For training the generator and discriminator of the StyleGAN2 we use an Adam optimizer with a learning rate of $0.0025$, $\beta_1$ as $0.0$, and $\beta_2$ as $0.99$. 

\textit{Encoder Training:}
We use a ResNet-34 (a stack of 34 residual blocks)~\cite{he2016deep} backbone as the architecture for our GAN Encoder network. We use an amplitude thresholding of -17dB for Water and -25dB for the Greatest Hits. That is, we mask the frequency components with magnitude below -17 or -25dB for the respective datasets. We use an Adam optimizer to train the Encoder with a learning rate of $0.00001$, $\beta_1$ as $0.5$, and $\beta_2$ as $0.99$.

\textit{Gaver Sound Synthesis:}
In all our experiments, we use 10 synthetic Gaver examples (5 per semantic attribute cluster) to generate the guidance vectors for controllable generation. We outline a cluster-based analysis for real and synthetic sounds using UMAP~\cite{mcinnes2018umap} visualizations on our supplementary webpage.

\subsection{Evaluation metric}
For audio quality, we utilize the \textit{Fréchet Audio Distance}~\cite{kilgour2019frechet}(FAD) metric. FAD is the distance between the distribution of the embeddings of real and synthesized audio data extracted from a pre-trained VGGish model. We utilize this metric to evaluate the quality of sounds generated by inverting the synthetic Gaver sounds as well as real-world sounds from the latent space of the GAN.

To evaluate the effectiveness of our method in changing a semantic attribute of a texture, we perform \textit{rescoring analysis}. By \textit{rescoring}, we mean the change in accuracy scores reported by an attribute classifier before and after the change in the semantic attribute on a sound. For this, we train an attribute presence or absence classifier based on a Dense Convolutional Network (or DenseNet) architecture~\cite{huang2017densely}. Previously,~\cite{palanisamy2020rethinking} showed that an ImageNet pre-trained model fine-tuned for audio datasets can be used to achieve state-of-the-art results in environmental sound classification tasks. We adapt the classifier from~\cite{palanisamy2020rethinking} using a DenseNet architecture with ImageNet pre-training and fine-tune it for our attribute classification task. Please see our supplementary webpage for classifier architecture and training details. 

We begin our evaluation by first manipulating an attribute on some randomly generated sounds. We then record how the attribute classifier score changes for those sounds before and after the manipulation. Further, we evaluate if the attribute change occurs without modifying other attributes of the sound. For instance, when editing \textbf{\textit{Brightness}}, we first analyze if the intended attribute of brightness changes. We then analyze if other attributes such as \textbf{\textit{Rate}} changes with it. As our datasets are unlabelled, to train the \textit{rescoring analysis} classifier, we manually curate and label a small subset of sounds. To do this, we selected approximately 250 samples of 2-second sounds for each semantic attribute under consideration. This manual curation involved visually analyzing the video and auditioning the associated sounds to detect the semantic attribute. For more details on the dataset curation please see our webpage. Note that this curated dataset is only used for quantitative analysis and not used to train our GAN or Encoder models.

\begin{table*}[t]
    \caption{Ablation Studies}
    \label{tab:ablation_studies}
    \centering
    \begin{tabular}{|p{4.3cm}|p{1.4cm}|p{1.4cm}|p{1.4cm}|p{1.4cm}|p{1.4cm}|p{1.4cm}|p{1.4cm}|}
    \hline
    &\multicolumn{6}{c|}{Greatest Hits}&Water\\
    \hline
    &\multicolumn{2}{c|}{Brightness}&\multicolumn{2}{c|}{Rate}&\multicolumn{2}{c|}{Impact Type}&Fill-Level\\
    \hline
    &Acc.($\uparrow$)&Avg. Chng Others ($\downarrow$)&Acc.($\uparrow$)&Avg. Chng Others ($\downarrow$)&Acc.($\uparrow$)&Avg. Chng Others ($\downarrow$)&Acc.($\uparrow$)\\
    \hline
    EBF: MSE+LPIPS~\cite{zhang2018unreasonable}&$0.53\pm\scriptscriptstyle{0.08}$&$0.24\pm\scriptscriptstyle{0.04}$&$0.64\pm\scriptscriptstyle{0.08}$&$0.35\pm\scriptscriptstyle{0.07}$&$0.69\pm\scriptscriptstyle{0.07}$&${0.20}\pm\scriptscriptstyle{0.04}$&$0.60\pm\scriptscriptstyle{0.1}$\\
    \hline
    EBF: MSE&$0.71\pm\scriptscriptstyle{0.06}$&${0.14}\pm\scriptscriptstyle{0.03}$&$0.88\pm\scriptscriptstyle{0.06}$&${0.30}\pm\scriptscriptstyle{0.07}$&$\boldsymbol{0.81}\pm\scriptscriptstyle{0.04}$&$0.24\pm\scriptscriptstyle{0.4}$&$0.27\pm\scriptscriptstyle{0.1}$\\
    \hline
    EBF: MSE+Thresholding $^{\dagger}$ &$\boldsymbol{0.82}\pm\scriptscriptstyle{0.05}$&$0.21\pm\scriptscriptstyle{0.06}$&$\boldsymbol{0.89\pm\scriptscriptstyle{0.06}}$&$0.35\pm\scriptscriptstyle{0.1}$&$0.80\pm\scriptscriptstyle{0.06}$&$0.27\pm\scriptscriptstyle{0.06}$&$\boldsymbol{0.97}\pm\scriptscriptstyle{0.04}$\\
    \hline
    \end{tabular}
\end{table*}

\subsection{Baseline Selection}

\begin{table*}[t]
    \caption{FAD Scores for GAN generated sounds and Encoder reconstructions}
    \label{tab:fad_scores}
    \centering
    \begin{tabular}{|p{4.3cm}|p{0.5cm}|p{1.4cm}|p{1.4cm}|p{0.5cm}|p{1.4cm}|p{1.4cm}|p{1.4cm}|}
    \hline
    &\multicolumn{3}{c|}{Greatest Hits}&\multicolumn{3}{c|}{Water}\\
    \hline
    &GAN&GAN Recon.$(\downarrow)$&Gaver Recon.$(\downarrow)$&GAN&GAN Recon.$(\downarrow)$&Gaver Recon.$(\downarrow)$\\
    \hline
    EBF: MSE+LPIPS~\cite{zhang2018unreasonable}&\multirow{ 4}{*}{$0.6$}&$1.12$&$4.40$&\multirow{ 4}{*}{1.17}&$1.92$&$9.45$\\\cline{1-1}\cline{3-4}\cline{6-7}
    EBF: MSE&&$\boldsymbol{0.72}$&$4.61$&&$1.59$&$11.77$\\\cline{1-1}\cline{3-4}\cline{6-7}
    EBF: MSE+Thresholding $^{\dagger}$&&$2.83$&$\boldsymbol{4.16}$&&$\boldsymbol{1.42}$&$\boldsymbol{7.92}$\\
    \hline
    \end{tabular}
\end{table*}
We evaluate our method's effectiveness in finding user-defined attribute guidance vectors in the latent space of the GAN by comparing it with an unsupervised method for latent semantic discovery called closed-form Semantic Factorization (SeFa)~\cite{shen2021closed}. SeFa decomposes the pre-trained weights of a GAN to find statistically significant vectors for guided generation. Although SeFa is relatively under-studied in the domain of audio, we use it as a baseline for comparison because, like our method, SeFa works on unconditionally trained GANs. Given the novelty of our task in deriving guidance vectors in a post-hoc fashion, to the best of our knowledge, the SeFa method is the state-of-the-art method in this regard. We thus use it for comparison.
\subsection{Experimental Details}
We first conduct ablation studies to understand the effects of individual components of the loss functions outlined in section~\ref{subsec:gan-inversion}. We report this analysis using both \textit{rescoring analysis} and \textit{FAD} scores. Next, we study the impact of the change of an attribute on other attributes under consideration. We then compare our method (EBF) with the SeFa method as a baseline. Finally, we qualitatively study the effectiveness of our method by conducting listening tests. 
Figure~\ref{fig:teaser} shows some spectrogram examples of guided generation using our method. The standard error of means in all tables in this section was reported by bootstrapping the samples over 100 iterations.

\subsubsection{Ablation Studies}\label{subsubsec:ablations}
We conduct three types of ablation studies in our paper - (1) to study the effect of different components of the loss function on the Encoder, (2) to study the effect of the number of synthetic samples needed to create a semantic cluster, and (3) to study the effect of the magnitude of scalar $\alpha$ in equation~\ref{eq:direction_edit}.

\textit{Ablating Encoder Loss Components}: We first study the effect of using LPIPS and MSE loss terms with and without thresholding while training the Encoder. Table~\ref{tab:ablation_studies} shows the \textit{rescoring} accuracy scores for attribute changes for each type of Encoder. $(\uparrow)$ indicates that higher values are better. For each attribute change, we report the accuracy for the main attribute as well as the average change reported in other attributes. Ideally, we would like the main attribute accuracy to be high and the change in other attributes to be low. 
For the Greatest Hits dataset, we find that by using MSE and MSE+Thresholding, the system outperforms the one with MSE+LPIPS loss for all attributes. For Water, using an Encoder with MSE and thresholding works best. 

Table~\ref{tab:fad_scores} shows the FAD scores for GAN-generated sounds (column called GAN) and Encoder reconstructions for each type of Encoder for both GAN-generated sounds as well as synthetic Gaver sounds. The FAD Scores were computed based on 10,000 randomly generated samples in comparison with the entire training set. We find that the encoder trained using MSE only or MSE+thesholding outperforms MSE+LPIPS in terms of the quality of the generated audio (FAD scores). Thus, based on this table and Table~\ref{tab:ablation_studies}, we choose the Encoder with MSE+Thresholding (qualified with a $\dagger$ in Tables~\ref{tab:ablation_studies} and ~\ref{tab:fad_scores}) as the best-performing Encoder for both datasets for the remainder of the paper. 

\textit{Ablating the Number of Gaver Samples}: Next, we study the effect of the number of Gaver samples used to find the guidance vectors for different attributes. We derive guidance using different $N$, starting with $N$=1 to $N$=5. We observe that as $N$ increases, the effectiveness of the directional vector edits also increases. Also, such edits preserve other unedited attributes better with higher $N$. The samples with different $N$'s can be auditioned on our supplementary webpage.

\textit{Ablating the effect of the scalar value $\alpha$}: In equation~\ref{eq:direction_edit}, the scalar value $\alpha$ governs the magnitude of the edit performed on the sample $\mathbf{w}$ using the semantic attribute direction vector $\mathbf{d}$. In all our experiments, the value of $\alpha$ is in between $[0,1]$. Also, all examples on our supplementary webpage edit $\mathbf{w}$ in linear steps until $\alpha = 1$. In this section, we qualitatively study the effect of using a value $\alpha > 1$, i.e., extrapolating the semantic edits beyond the magnitude of the difference vector $\mathbf{d}$ (or beyond the selected prototype in the latent manifold). The samples from different $\alpha$'s can be found on our webpage. We observe that, for all attributes, for $\alpha >= 3$, the edited $\mathbf{w}$ vectors escape the latent $\mathcal{W}$ manifold and generate noisy or unintelligible samples.

\subsubsection{Baseline Comparison}
\begin{table*}[t]
    \caption{Comparison with Baseline}
    \label{tab:baseline_comparison}
    \centering
    \begin{tabular}{|p{4cm}|p{1.4cm}|p{1.4cm}|p{1.4cm}|p{1.4cm}|p{1.4cm}|p{1.4cm}|p{1.4cm}|}
    \hline
    &\multicolumn{6}{c|}{Greatest Hits}&Water\\
    \hline
    &\multicolumn{2}{c|}{Brightness}&\multicolumn{2}{c|}{Rate}&\multicolumn{2}{c|}{Impact Type}&Fill-Level\\
    \hline
    &Acc.($\uparrow$)&Avg. Chng Others ($\downarrow$)&Acc.($\uparrow$)&Avg. Chng Others ($\downarrow$)&Acc.($\uparrow$)&Avg. Chng Others ($\downarrow$)&Acc.($\uparrow$)\\
    \hline
    SeFa~\cite{shen2021closed}&$0.49\pm\scriptscriptstyle{0.11}$&$0.19\pm\scriptscriptstyle{0.12}$&$0.45\pm\scriptscriptstyle{0.12}$&$0.29\pm\scriptscriptstyle{0.14}$&$0.42\pm\scriptscriptstyle{0.15}$&$0.31\pm\scriptscriptstyle{0.09}$&$0.92\pm\scriptscriptstyle{0.09}$\\
    \hline
   EBF: MSE+Thresholding $^{\dagger}$ &$\boldsymbol{0.82}\pm\scriptscriptstyle{0.05}$&$0.21\pm\scriptscriptstyle{0.06}$&$\boldsymbol{0.89\pm\scriptscriptstyle{0.06}}$&$0.35\pm\scriptscriptstyle{0.10}$&$\boldsymbol{0.80}\pm\scriptscriptstyle{0.06}$&$0.27\pm\scriptscriptstyle{0.06}$&$\boldsymbol{0.97}\pm\scriptscriptstyle{0.04}$\\
    \hline
    \end{tabular}
\end{table*}

\begin{table}[t]
\caption{Pairwise rescoring for Greatest Hits (EBF)}
\label{tab:rescoring_our_method}
\centering
\renewcommand{\arraystretch}{1.7} 
\begin{tabular}{cccc}
&Brightness$(\uparrow)$&Rate$(\uparrow)$&Impact Type$(\uparrow)$\\
Brightness &\cellcolor[rgb]{0.98,0.4,0.4}$\boldsymbol{0.82}\pm0.06$&\cellcolor[rgb]{0.98, 0.95, 0.95}$0.06\pm0.04$&\cellcolor[rgb]{0.98, 0.6, 0.6}$0.40\pm0.07$\\
Rate&\cellcolor[rgb]{.95, .60, .60}$0.40\pm0.09$&\cellcolor[rgb]{0.98,0.4,0.4}$\boldsymbol{0.89}\pm0.06$&\cellcolor[rgb]{.98, .62, .62}$0.38\pm0.09$\\
Impact Type&\cellcolor[rgb]{.95, .75, .75}$0.35\pm0.07$&\cellcolor[rgb]{0.98, 0.81, 0.81}$0.19\pm0.03$&\cellcolor[rgb]{0.98,0.4,0.4}$\boldsymbol{0.80}\pm0.05$\\
\end{tabular}
\bigskip
  \caption{Pairwise rescoring for Greatest Hits (SeFa)}
  \label{tab:rescoring_sefa_method}
  \renewcommand{\arraystretch}{1.7} 
  \begin{tabular}{cccc}
    &Brightness$(\uparrow)$&Rate$(\uparrow)$&Impact Type$(\uparrow)$\\
Dimension 0&\cellcolor[rgb]{0.98, 0.9, 0.9}$0.10\pm0.07$&\cellcolor[rgb]{0.98, 0.93, 0.93}$0.07\pm0.06$&\cellcolor[rgb]{.98, .82, .82}$0.18\pm0.13$\\
Dimension 1&\cellcolor[rgb]{.98, .70, .70}$0.30\pm0.11$&\cellcolor[rgb]{0.98,0.55,0.55}$\boldsymbol{0.45}\pm0.12$&\cellcolor[rgb]{.98, .72, .72}$0.28\pm0.16$\\
Dimension 2&\cellcolor[rgb]{.98, .69, .69}$0.31\pm0.11$&\cellcolor[rgb]{0.98, 0.91, 0.91}$0.09\pm0.06$&\cellcolor[rgb]{0.98,0.58,0.58}$\boldsymbol{0.42}\pm0.15^{\boldsymbol{*}}$\\
Dimension 3&\cellcolor[rgb]{0.98,0.51,0.51}$\boldsymbol{0.49}\pm0.12$&\cellcolor[rgb]{0.98, 0.91, 0.91}$0.09\pm0.06$&\cellcolor[rgb]{.98, .70, .70}$0.30\pm0.16$\\
Dimension 4&\cellcolor[rgb]{.98, .88, .88}$0.12\pm0.08$&\cellcolor[rgb]{.98, .86, .86}$0.14\pm0.08$&\cellcolor[rgb]{.98, .83, .83}$0.17\pm0.12$\\
Dimension 5&\cellcolor[rgb]{.98, .69, .69}$0.31\pm0.11$&\cellcolor[rgb]{0.98, 0.9, 0.9}$0.10\pm0.06$&\cellcolor[rgb]{.98, .70, .70}$0.30\pm0.14$\\
Dimension 6&\cellcolor[rgb]{.98, .68, .68}$0.32\pm0.11$&\cellcolor[rgb]{.98, .92, .92}$0.14\pm0.08$&\cellcolor[rgb]{.98, .81, .81}$0.19\pm0.13$\\
Dimension 7&\cellcolor[rgb]{.98, .86, .86}$0.14\pm0.08$&\cellcolor[rgb]{0.98, 0.9, 0.9}$0.10\pm0.06$&\cellcolor[rgb]{.96, .58, .58}$\boldsymbol{0.38}\pm0.16^{\boldsymbol{*}}$\\
Dimension 8&\cellcolor[rgb]{.98, .82, .82}$0.18\pm0.09$&\cellcolor[rgb]{0.98, 0.91, 0.91}$0.09\pm0.07$&\cellcolor[rgb]{.98, .72, .72}$0.28\pm0.16$\\
Dimension 9&\cellcolor[rgb]{.98, .66, .66}$0.34\pm0.10$&\cellcolor[rgb]{.98, .89, .89}$0.11\pm0.07$&\cellcolor[rgb]{.98, .8, .8}$0.20\pm0.13$\\

\end{tabular}
\end{table}

\begin{table}[t]
\caption{Pairwise rescoring for Water (EBF and SeFa)}
\parbox{.45\linewidth}{
\label{tab:rescoring_water-ours}
\centering
\renewcommand{\arraystretch}{1.7} 
\begin{tabular}{cc}
\textbf{EBF$^{\dagger}$}&Fill-Level$(\uparrow)$\\
Fill-Level&\cellcolor[rgb]{0.98,0.35,0.35}$\boldsymbol{0.97}\pm0.04$\\
\end{tabular}
}
\parbox{.45\linewidth}{
\label{tab:rescoring_water-sefa}
\centering
\renewcommand{\arraystretch}{1.7} 
\begin{tabular}{cc}
\textbf{SeFa}&Fill-Level$(\uparrow)$\\
Dim 0&\cellcolor[rgb]{0.98, 0.86, 0.86}$0.14\pm0.11$\\
Dim 1&\cellcolor[rgb]{0.98,0.35,0.35}$\boldsymbol{0.92}\pm0.09$\\
Dim 2&\cellcolor[rgb]{.98, .73, .73}$0.27\pm0.16$\\
\end{tabular}
}
\end{table}

Table~\ref{tab:baseline_comparison} reports the \textit{rescoring analysis} for each attribute using our method in comparison with SeFa. We report the score for change in the main intended attribute being edited and the average change in other attributes. $(\uparrow)$ indicates higher values are better. For both datasets, our method reports better accuracies for change in the main attribute than SeFa. 

We further report pairwise attribute edit comparisons to study the effect of change in one attribute individually on every other attribute. Tables~\ref{tab:rescoring_our_method} and \ref{tab:rescoring_sefa_method} show this for the Greatest Hits dataset and Table~\ref{tab:rescoring_water-ours} for the Water dataset. For SeFa, since we do not know which vector (of the $\delta_w$=128 dimensions) edits a specific attribute, we report scores for edits performed by the top 10 vectors with the highest singular values (top 10 for Greatest Hits and top 3 for Water Filling) for comparison in the table. $(\uparrow)$ indicates higher values are better and scores highlighted with ''*'' indicates no significant differences ($p > 0.05$). Each row indicates a semantic attribute manipulation using a specific guidance vector and each column evaluates how the scores changed for that attribute. The darkened cells in the table indicate dimensions with the highest score for a semantic attribute (in that column).

For both datasets, our method reports a significant change in the main attribute being manipulated. Further, we analyze if each dimension or direction vector from both methods manipulates only a single attribute. For this, we perform a two-way t-test for the scores between any two SeFa dimensions. We particularly notice that for SeFa, the semantic attribute of \textit{\textbf{Impact Type}} is affected by at least two dimension vectors, namely Dimension 2 and Dimension 7 in Table~\ref{tab:rescoring_sefa_method}. This implies that methods such as SeFa may not always guarantee one-to-one correspondence between statistically found vectors for guidance and the semantic attributes of interest. Furthermore, the first dimension associated with the largest singular value extracted using SeFa does not correlate with any of the main perceptually varying attributes in both datasets. This implies that such automated methods do not always guarantee to find vectors that control perceptually relevant attributes in the latent space of a generative model for audio.

\subsubsection{Listening tests}
\begin{table}[t]
\caption{Listening Test Results}
\label{tab:listening_test_results}
\centering
\renewcommand{\arraystretch}{1.3} 
\begin{tabular}{|p{0.5cm}|p{1.5cm}|p{1.6cm}|p{1.45cm}|p{1.45cm}|}
\hline
&Water&\multicolumn{3}{c|}{Greatest Hits}\\
\hline
&Fill Level$(\uparrow)$&Brightness$(\uparrow)$&Rate$(\uparrow)$&Impact Type$(\uparrow)$\\
\hline
SeFa&$0.47\pm0.03$&$0.75\pm0.03^*$&$0.58\pm0.04$&$0.51\pm0.04$\\
\hline
EBF$^{\dagger}$&$\mathbf{0.55}\pm\mathbf{0.03}$&$0.75\pm0.04^*$&$\mathbf{0.68}\pm\mathbf{0.04}$&$\mathbf{0.67}\pm\mathbf{0.05}$\\
\hline
\end{tabular}
\end{table}
We recruited 20 participants on Amazon's Mechanical Turk (AMT) to evaluate the sounds modified by using both methods. Only participants with more than 95\% approval rate on their previous tasks on AMT across at least 1000 completed tasks were allowed to attempt our listening test. Before attempting our listening test, participants underwent a hearing screening designed for crowdsourced platforms based on~\cite{cartwright2016fast}. The participants were requested to sit in a quiet place and use a pair of headphones for the duration of the test. During the hearing screening, the participants were presented with two audio samples. Each sample contained different tones generated at random frequencies between $55Hz$ and $10kHz$. They were asked to count the number of tones in each audio sample. Participants who completed the screening by correctly estimating the number of tones were allowed to attempt our listening test. The audio samples in the hearing screening ensured that the participants were of normal hearing, were using a pair of headphones, and were in a quiet environment when attempting the listening test. 

We created the audio samples for our main listening test by randomly sampling from the StyleGAN and then editing each sample using the direction vectors using both methods. For the Greatest Hits dataset, we randomly sampled 20 sounds from the StyleGAN2's latent space and modified them using vectors derived using our method for \textbf{\textit{Brightness}}, \textbf{\textit{Impact Type}} and \textbf{\textit{Rate}}. For SeFa, we used the vectors with the highest rescoring accuracy from Table~\ref{tab:rescoring_sefa_method} to manipulate the samples. We developed a listening test interface to evaluate our attribute edits. The participants were presented with the unmodified original reference sound as well as the manipulated samples. They were asked to evaluate if the two samples differed in the 3 attributes. For the Water dataset, we randomly sampled the latent space 10 times and modified the samples using vectors for \textbf{\textit{Fill-Level}}. As the \textbf{\textit{Fill-Level}} for Water varies continuously, we wanted to evaluate if manipulating the sound samples sequentially and linearly using both methods preserves the interim \textbf{\textit{Fill-Level}}s (such as when the bucket is empty, quarter or half full, etc.). To do this, we use the rank-ordering interfaces outlined in~\cite{10.1145/3581641.3584083} to measure the perceptual linearity of linearly manipulating the sample using the guidance vector for \textbf{\textit{Fill-Level}}. The interfaces for the listening tests can be viewed on our supplementary webpage.

We use accuracy scores to evaluate our listening tests, with `accuracy' formulated as the fraction of the listening test trials where participants correctly selected the attribute being manipulated for a sample in comparison to a reference. Table~\ref{tab:listening_test_results} shows the scores from our listening tests for both datasets and their respective attributes. $(\uparrow)$ indicates that higher values are better and scores highlighted with ``*'' indicates no significant differences $(p>0.05)$. For Water, participants were able to perceptually rank-order the water-filling sounds in increasing order of \textbf{\textit{Fill-Level}} significantly better when using our method. For the Greatest Hits dataset, participants found our method to perform significantly better while manipulating the sounds for \textbf{\textit{Rate}} and \textbf{\textit{Impact Type}}. However, for the attribute of \textbf{\textit{Brightness}}, participants found both methods to perform equally well. By qualitatively listening and comparing the brightness samples generated by the algorithm, we find that samples generated using our method cover a wider range of brightness than SeFa (visit the supplementary webpage for examples).

\section{Application: Selective Semantic Attribute Transfer}
In this section, we demonstrate the simplicity of extending our framework to applications other than performing semantic edits of textures. The prototypes and guidance vectors derived from our method can be used to support applications such as selective semantic attribute transfer. This task is inspired by image editing applications such as Photoshop, where a user can select an object and transfer its color to another object. We envision a selective attribute transfer tool where the prototype and guidance vectors guide the process of selecting an attribute from a reference sample and transferring it to another sample.

\begin{figure}[t]
     \centering
     \includegraphics[width=0.4\textwidth]{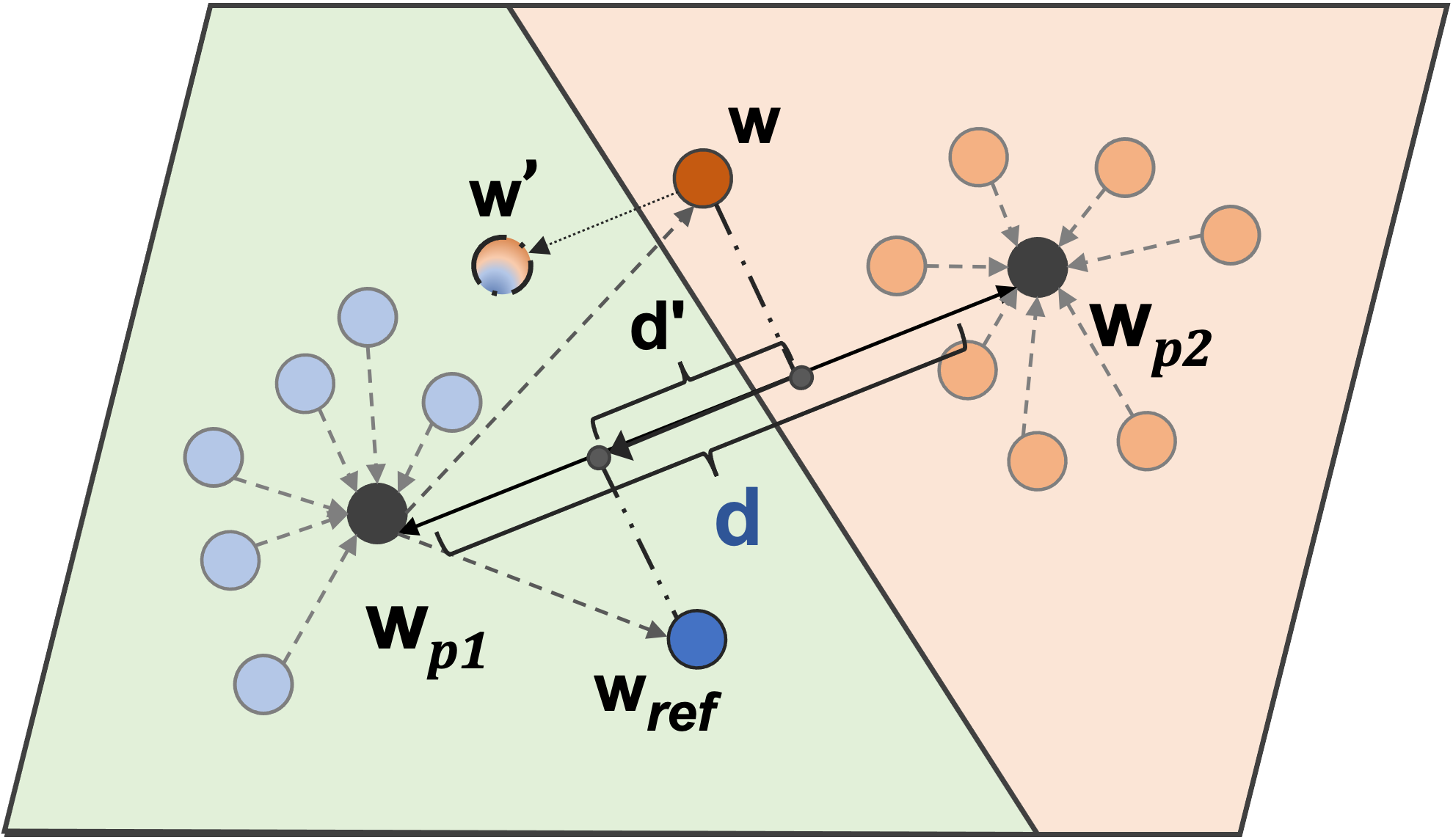}
     \caption{Semantic attribute transfer from a reference sample $\mathbf{w_{ref}}$ to a target $\mathbf{w}$, with direction vector $\mathbf{w_{p1}}\rightarrow\mathbf{w_{p2}}$ representing, say an  increasing level of ``Brightness''. Both $\mathbf{w_{ref}}$ and $\mathbf{w}$ are projected onto to the direction vector $\mathbf{d}$. The difference vector $\mathbf{d'}$ is used to selectively edit $\mathbf{w}$ to generate $\mathbf{w'}$. $\mathbf{w'}$ will have the same brightness relationship to $\mathbf{w}$ as $\mathbf{w_{ref}}$.}\label{fig:concept-prototypes-schematic-2}
\end{figure}

Figure~\ref{fig:concept-prototypes-schematic-2} shows a diagram outlining the approach. Say we have a reference sample embedding $\mathbf{w_{ref}}$ and a target sample embedding $\mathbf{w}$, and we want to selectively transfer the attribute of \textbf{\textit{Brightness}} from the reference $\mathbf{w_{ref}}$ to $\mathbf{w}$. To do this, we first project both $\mathbf{w_{ref}}$ and $\mathbf{w}$ onto the attribute guidance vector $\mathbf{d}$ between $\mathbf{w_{p1}}$ and $\mathbf{w_{p2}}$. We then edit $\mathbf{w}$ in the direction of the difference between the projections, namely $\mathbf{d'}$, to create $\mathbf{w'}$. This method not only transfers the \textbf{\textit{Brightness}} attribute from $\mathbf{w_{ref}}$ to $\mathbf{w'}$ but also preserves the other unmodified semantic attributes of $\mathbf{w}$ as well as its original structure (position or location of the impact events along the time axis). 
A formal outline of this algorithm, as well as some results from selectively transferring individual attributes such as \textbf{\textit{Brightness}} onto a target sample, can be found on our webpage. 
 
\section{Limitations and Future Work}
In this section, we explore a few limitations of our method that surfaced during our exploration of user-defined semantic attribute guidance through the latent space of a StyleGAN. We outline future work on how to improve our method's effectiveness, its applicability to other sound types, and the use of other audio querying mechanisms.

\subsubsection{Constraining traversal to the latent manifold for $\alpha > 1$}
In section~\ref{subsubsec:ablations}, we qualitatively study the effect of using $\alpha > 1$ to perform edits. As seen in equation~\ref{eq:direction_edit}, our method assists in a linear traversal of the latent space using the computed direction vector to perform semantic edits on any randomly generated samples. For higher values of $\alpha$, this linear way of traversing the latent space may result in $\mathbf{w}$ vectors falling outside of the latent manifold. Such edited vectors may result in the generation of noisy or unintelligible samples. Other traversal methods may accommodate the latent space's local geometry such that local edits do not escape the latent manifold. One approach is to use more than two prototypes, supported by ``in-between'' examples, to improve the robustness of our framework. Another approach is to investigate the use of incremental non-linear traversal methods (such as in~\cite{choi2022do}). With such traversal methods, the guidance can be directed to stay within the manifold by incrementally guiding the generation using a sequence of consecutive $\mathbf{d}$ vectors. However, it should be noted that such methods will increase the number of computations to be performed during edits or the number of synthetic samples that will need to be manually created. 

\subsubsection{Manual curation of samples}
Although our method has been more effective than algorithms such as SeFa at modifying the user-defined semantic attributes on a texture, some manual curation of synthetic samples is needed to find the relevant guidance vectors. On the other hand, algorithms such as SeFa are automatic and can be applied to any pre-trained GAN without any manual intervention. Thus, in our future work, we will explore the potential of combining SeFa's ability to automatically discover vectors for attribute manipulation with our method to improve the accuracy of editing the semantic attributes.

\subsubsection{Querying GANs using out-of-distribution sounds}
The Encoder outlined in section~\ref{subsec:gan-inversion} is trained on masked and amplitude thresholded versions of the real-world training data. This approach assists in projecting any out-of-training-distribution sounds, such as the parametrically synthesized sounds in our case, to a reasonable part of the latent space, even though the Encoder is not directly trained on such synthetic sounds. Such an Encoder can be extended to querying the StyleGAN's latent space using other out-of-distribution sounds such as the sound generated vocally by users (i.e., query-by-humming approaches). A productive avenue for future work will be in further studying the applicability or limitations of our framework in conjunction with vocal queries to perceptually guide audio texture generation.

\subsubsection{Applicability to music and other environmental sounds}
While we demonstrate the efficacy of our method for perceptually guiding the generation of audio textures, a potential limitation of extending this approach to other sound types is in the parametric synthesizer (in section~\ref{subsec:analysis-by-synthesis-approach}). Our current parametric synthesis technique is limited by its ability to model sounds based on object resonances or physical parameters of the interacting objects. Newer approaches need to be developed to approximate the synthetic sound queries needed for navigating the latent space of other sound types, such as timbres from musical instruments as well as speech and other environmental sounds (e.g., footsteps or machine sounds). A potential avenue for such procedural synthesis models can be found in~\cite{wyse2022syntex}.

\subsubsection{Approaching semantic edits using text-to-audio models}
Recently, text-to-audio models, which rely on the supervision of text captions during training, have made significant progress in demonstrating controllability during the generation of environmental sounds~\cite{kreuk2022audiogen,liu2023audioldm}. In our experiments, we restricted ourselves from performing a comparison with such models because the datasets we used in our study were unlabelled and were not associated with text captions needed to train text-to-audio models. Further, we refrained from using off-the-shelf text-to-audio models for comparison as their training data (such as Audioset~\cite{gemmeke2017audio}) significantly differed from the training data distribution under the purview of our work. Although we are unable to perform a systematic comparison of our method with text-to-audio models, on our supplementary webpage\footnote{\href{https://pkamath2.github.io/audio-guided-generation/\#t-to-a}{https://pkamath2.github.io/audio-guided-generation/\#t-to-a}} we demonstrate some text prompts that assist in achieving the semantic editing goals of our framework using text-to-audio models such as AudioGen~\cite{kreuk2022audiogen} and AudioLDM~\cite{liu2023audioldm}. For impact sounds, we designed prompts by describing the material properties of the impact surface, along with certain acoustic properties of the sound. Similarly, for water filling, we described the material properties of the container as well as the fill level of the water.

While it should be noted that well-engineered prompts would lead to better results, with the prompts we used (on our supplementary page), we observed that editing a prompt considerably changed not just the semantic attribute being edited but also other attributes of the sound. For instance, modifying an existing prompt by adding a Rate feature such as `fast' considerably changed other aspects of the sound, such as Brightness, and also removed the `long sustain' from the originally prompted sound. This could be because in text-to-audio models text prompts could be entangled with multiple semantic attributes of the sounds. This observation warrants further systematic experimentation to study using text to semantically edit sounds in comparison (or in conjunction) with sound-based frameworks such as ours.

\section{Conclusion}
In this paper, we propose an audio example-based method to perceptually guide the generation of audio textures based on user-defined semantic attributes. By using a synthesizer to create a few examples, we can develop attribute guidance vectors in the latent space of a StyleGAN2 to controllably generate both impact sounds as well as continuously varying water-filling audio textures. We show the effectiveness of our method in providing linearly varying controls for texture generation using both objective metrics as well as perceptual listening tests. Furthermore, we demonstrate an application of our method to other signal-processing tasks, namely semantic attribute transfer.










\balance
\bibliographystyle{IEEEtran}
\bibliography{main}


\end{document}